\RequirePackage{lineno}
\documentclass{JINST}

\def\lsim{\mathrel{\rlap{\lower4pt\hbox{\hskip1pt$\sim$}}
    \raise1pt\hbox{$<$}}}
\def\gsim{\mathrel{\rlap{\lower4pt\hbox{\hskip1pt$\sim$}}
    \raise1pt\hbox{$>$}}}

\newcommand{\beq}{\begin{eqnarray}}
\newcommand{\eeq}{\end{eqnarray}}

\def\um{\mbox{m}}

\title{The MicroBooNE light collection system}

\author{
Teppei Katori$^a$ for the MicroBooNE collaboration\\ 
$^a$Physics Department, Massachusetts Institute of Technology, Cambridge, MA 02139\\
E-mail: \email{KATORI@FNAL.GOV}}

\abstract{MicroBooNE is a neutrino experiment 
located on axis in the Booster Neutrino Beamline (BNB), 
at Fermi National Accelerator Laboratory, 
scheduled to begin data collection in 2014. 
The MicroBooNE detector consists of two main components: 
a large liquid argon TPC, 
and a light collection system. 
Thirty-two 8-inch diameter cryogenic photomultiplier tubes (PMTs) 
will detect the scintillation light generated in the liquid argon. 
In this article, we describe the basic features of the system and current status of 
MicroBooNE light collection system.
}


\begin{document}

\section{PMT array}

MicroBooNE experiment is a part of the US liquid argon TPC (LArTPC) efforts~\cite{uB}, 
and it will provide critical information for future large LArTPC experiments, 
such as Long-Baseline Neutrino Experiment (LBNE)~\cite{LBNE}. 
MicroBooNE uses a low energy ($\sim$800~MeV) wide-band BNB neutrino beam 
to investigate the low energy event excesses observed in MiniBooNE~\cite{MB_osc}. 
The high statistics and high resolution data from MicroBooNE are expected to 
answer current mysteries in neutrino interactions, too~\cite{Teppei_NuInt11}.
The construction of the MicroBooNE detector is ongoing, 
and beam data taking is planned to start from 2014.   

\begin{figure}[b]
\centering
\includegraphics[width=1.00\textwidth]{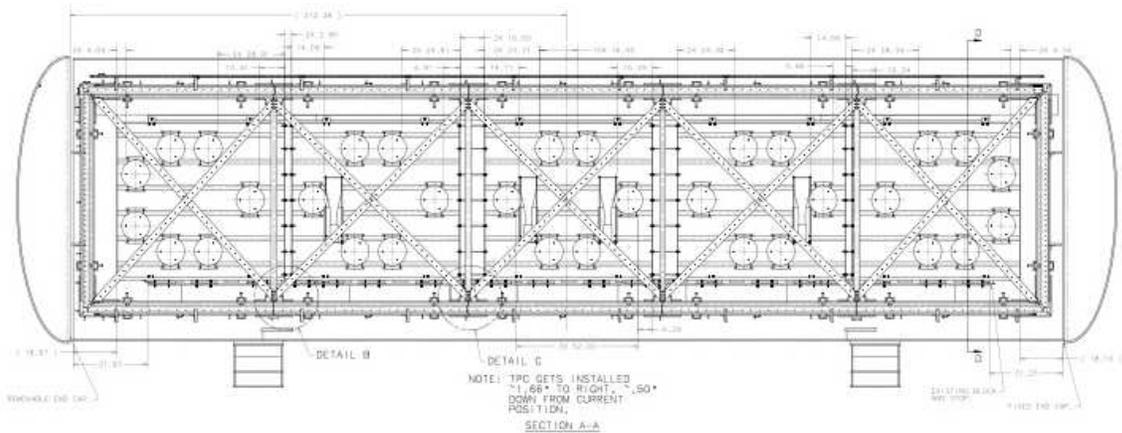}
\caption{\label{fig:rick}
A drawing to show the configuration of PMTs and light guide paddles 
in the MicroBooNE cryostat. 
Notice 32 PMTs (circles) and 4 light guide paddles 
(rectangular objects in middle row of central area) are located to avoid TPC frame ``crosses''.
}
\end{figure}

The 170 ton liquid argon volume in the MicroBooNE cryostat contains 
a $2.5\times2.4\times10.4~\um^3$ TPC drift volume. 
The basic strategy of the MicroBooNE PMT array largely follows 
the ICARUS T600 light collection system~\cite{ICARUS_detec}.
PMTs are located behind the wire planes to observe the argon scintillation light. 
The important feature of the light collection system is its ability to 
measure the event time at the $\sim$ns level.
Since prompt light from the liquid argon is emitted on a 3-6~ns time scale, 
detection of the scintillation light allows the LArTPC to be triggered, 
where drift of electrons take $\sim$ms. 

Figure~\ref{fig:rick} shows the PMT configuration. 
It also shows how light guide paddles are configured, described later. 
To achieve a uniform response, 
the PMTs are evenly distributed. 
The PMTs are additionally located to avoid the TPC frame which make several 
``crosses'' in front of the PMTs. 
Since the wire planes are located inside of the TPC frame, 
the distance from the collection wire plane 
to the PMT surface is about 10~cm. 

\subsection*{PMT unit}

\begin{figure}
\centering
\includegraphics[width=0.60\textwidth]{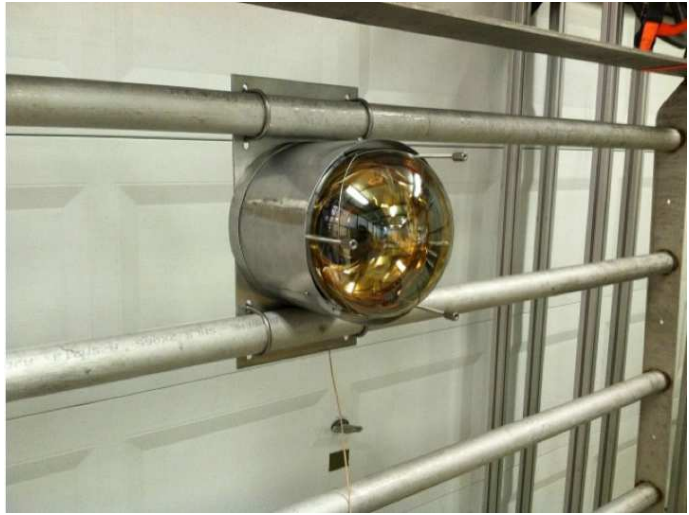}
\caption{\label{fig:PMTunit}
A picture of the mechanical model of the MicroBooNE PMT unit without the TPB plate in the test facility. 
}
\end{figure}

Each PMT is fit inside of the PMT mount, which is surrounded by the cryogenic magnetic shield. 
Finally, a wavelength shifter coated acrylic plate (TPB plate) is equipped in front of the PMT window. 
This makes one PMT unit, and 32 PMT units are mounted on the PMT rack. 
Figure~\ref{fig:PMTunit} shows the mechanical model of the MicroBooNE PMT unit without the TPB plate. 

\subsubsection*{PMT}

We use 32 Hamamatsu R5912-02mod 14 stage cryogenic 8-inch hemi-spherical PMTs. 
Large photocathode PMTs provide a cost-effective solution to maximise photocathode coverage; 
this design corresponds to a 0.9\% photocathode coverage of the TPC volume. 
It is known that the gain of a PMT drops in a cryogenic environment, 
therefore a high gain PMT, 
such as R5912-02mod, can compensate for the gain drop by increasing the high voltage value. 
We tested all PMTs used for the experiment, and confirmed this gain drop can be recovered 
easily by increasing the high voltage by 200 V. 
More details of the PMT base structure and 
the PMT test results can be found in elsewhere~\cite{PMTTest}.

\subsubsection*{PMT mount}

\begin{figure}
\centering
\includegraphics[width=0.60\textwidth]{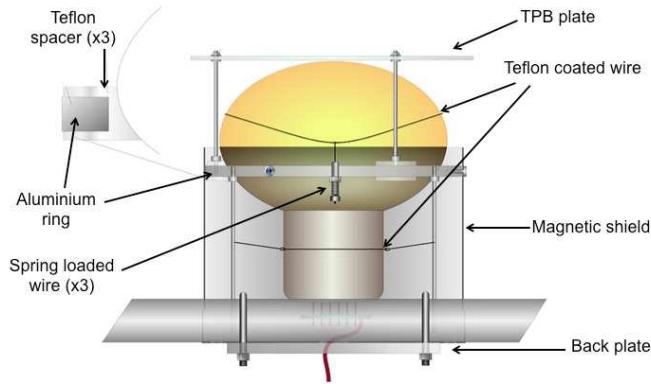}
\caption{\label{fig:PMTmount}
A schematic drawing of the PMT mount with a PMT. 
A wire ring is pulled down by 3 spring loaded wires to an aluminium ring. 
Direct contact of the PMT to the aluminium ring is avoided by 
Teflon blocks. The magnetic shield and the TPB plate are fixed 
on to the PMT mount. 
}
\end{figure}

Figure~\ref{fig:PMTmount} shows a schematic drawing of the PMT mount. 
A PMT sits on an aluminium ring, where direct contact of the PMT and the ring 
is prevented by 3 Teflon spacers. 
A Teflon coated wires run across the window. 
It makes a small shadow on the photo-cathode but the effect is negligible. 
Then this wire is pulled to the ring side, by 3 spring loaded wires. 
Extensive tests show this design secure the PMT from the forces acting on 
both downward (by the gravity, during the installation in air) 
and upward (by the buoyancy, during the operation in LAr). 
Notice all metal-glass contacts are avoided by Teflon. 
On the other hand, the thermal contraction of aluminium and Teflon is 
managed by the metal springs. 

In front of the PMT window, a TPB plate (described later) is supported by 3 posts. 
There are another 3 posts, running from the other side of the ring, 
which support the PMT mount and they are fixed on the stainless back plate. 
This back plate support all forces acting on the mount. 
It is attached on the PMT rack by U-bolts. 
There is a Teflon coated wire ring connected to posts, 
which loosely grabs the PMT stem to prevent the large tilt.
 
\subsubsection*{Cryogenic magnetic field shield}

A standard 304L stainless steel vessel, such as the MicroBooNE cryostat, is non-magnetic. 
However, it is shown that a hemi-spherical PMT is rather vulnerable to 
weak ambient magnetic fields~\cite{DCshield} 
such as the earth's magnetic field ($\sim$300 mG), 
which deflects the electron trajectory. 
This effect loses the PMT efficiency up to 50\%.  
Therefore, all PMTs are installed inside of a ``coffee can'' shaped cryogenic magnetic shield~\cite{AK4}. 
Note permeability generally decreases with temperature, however, 
specially annealed cryogenic magnetic shields have higher permeability in the cryogenic environment.  
The cryogenic magnetic shield has been tested in a specially made PMT test stand, 
by tilting PMT unit to see the effect of the earth magnetic field with and without 
the shield in air and liquid nitrogen. 
Without a magnetic shield, 
the photon detection efficiency is sensitive with the orientation of the PMT.  
However, it is confirmed that these magnetic shields can prevent that both in air and liquid nitrogen. 
From these tests, it is decided to cover the PMT by the cryogenic magnetic shield,  
from the equator of the window glass to the end of the stem region, 
including the PMT base~\cite{uB_BShield}. 
 
\subsubsection*{TPB plate}

Since the scintillation light from argon is 128~nm where a borosilicate window has zero transparency, 
we need a wavelength shifter to detect the scintillation light by the PMTs. 
Tetra phenyl Butadiene (TPB) is widely used in this community. 
Although there are several choices how to deposit TPB, 
we decided to paint an acrylic plate equipped in front of the PMT 
using a TPB-polystyrene mixture (50\%-50\% by mass)~\cite{benchmark}.  
In this way, the TPB plate is easy to produce in the laboratory, 
and R\&D and the quality control of the wavelength shifter are more simple. 
The quality control of the TPB plate is important 
since TPB is known to be degraded by exposure to UV light~\cite{Christie} 
through the production of benzophenone~\cite{benzophenone}. 
Consequently, the production of TPB plates will be the last stage of installation of the light collection system.


\section{Light guide paddle}

TPB-coated light guide have been developed~\cite{lightguide} as a candidate light collection detector 
for the LBNE experiment. 
The main advantage is that such light guide can be located between a narrow region of 2 collection wire planes. 
The 4 light guides will be installed in the MicroBooNE cryostat for R\&D.  
Fig.~\ref{fig:rick} shows where these 4 light guide paddles are located. 
Each paddle is made from acrylic, coated with a TPB-acrylic mixture, 
adiabatically bended to the window of a 2-inch R7725-mod cryogenic PMT. 
A comparison of PMT signals and light guide signals 
would explore the potential of such light collection method for future LArTPC detectors. 

\section{Installation}

The 32 PMT units and 4 light guide paddles are mounted on five separated PMT racks. 
A PMT rack has an ability to slide in the gap between the cryostat wall and the TPC frame, 
through a linear rail by an oil free Teflon coated linear bearing. 
In this way, we can install all PMTs after the installation of the TPC frame, 
to minimise the exposure of TPB plates 
to ambient light and moisture. 
In addition, the clean assembly tent surrounding the detector is UV protected.   
We are aiming to install the light collection system during summer 2013.
 

\acknowledgments

The author thanks to the organisers of LIDINE2013 for their invitation to the conference. 
The author also thanks to the strong support of Fermilab on this project. 
This work is supported by the National Science Foundation (NSF-PHY-0847843).

\end{document}